\documentclass[prd,twocolumn,groupedaddress,showpacs,nofootinbib]{revtex4}
\usepackage{graphicx}
\usepackage{dcolumn}
\usepackage{amssymb}
\usepackage{mathrsfs}
\usepackage{amsmath}
\usepackage{epsfig}
\usepackage[dvips]{color}
\usepackage{hhline}

\begin{document}

\title{From high-scale leptogenesis to low-scale one-loop neutrino mass generation }

\author{Hang Zhou}

\email{einsteinzh@sjtu.edu.cn}

\author{Pei-Hong Gu}

\email{peihong.gu@sjtu.edu.cn}

\affiliation{Department of Physics and Astronomy, Shanghai Jiao Tong University, 800 Dongchuan Road, Shanghai 200240, China}

\begin{abstract}

We show that a high-scale leptogenesis can be consistent with a low-scale one-loop neutrino mass generation. Our models are based on the $SU(3)_c^{}\times SU(2)_L^{}\times U(1)_Y^{}\times U(1)_{B-L}^{}$ gauge groups. Except a complex singlet scalar for the $U(1)_{B-L}^{}$ symmetry breaking, the other new scalars and fermions (one scalar doublet, two or more real scalar singlets/triplets and three right-handed neutrinos) are odd under an unbroken $Z_2^{}$ discrete symmetry. The real scalar decays can produce an asymmetry stored in the new scalar doublet which subsequently decays into the standard model lepton doublets and the right-handed neutrinos. The lepton asymmetry in the standard model leptons then can be partially converted to a baryon asymmetry by the sphaleron processes. By integrating out the heavy scalar singlets/triplets, we can realize an effective theory to radiatively generate the small neutrino masses at the TeV scale. Furthermore, the lightest right-handed neutrino can serve as a dark matter candidate.

\end{abstract}


\maketitle

\section{Introduction}

The phenomena of neutrino oscillations established by terrestrial experiments have indicated that three flavors of neutrinos should be massive and mixed \cite{patrignani2016}. This fact implies we need new physics beyond the standard model (SM) where the neutrinos are massless. On the other hand, cosmological observations have stringently constrained the neutrino masses should be extremely small \cite{patrignani2016}. Currently the seesaw  \cite{minkowski1977,yanagida1979,grs1979,ms1980,mw1980,sv1980,cl1980,lsw1981,ms1981,flhj1989,ma1998} extensions of the SM have become one of the most attractive schemes to generate the tiny neutrino masses because they can also accommodate a leptogenesis \cite{fy1986} mechanism to explain the baryon asymmetry in the universe.

In the usual seesaw models, some new particles with lepton-number-violating interactions can naturally suppress the neutrino masses by their heavy masses, meanwhile, the decays of these heavy particles can produce a lepton asymmetry to account for the cosmic baryon asymmetry in association with the sphaleron processes \cite{krs1985}. Clearly, the leptogenesis\cite{fy1986,lpy1986,luty1992,mz1992,fps1995,crv1996,pilaftsis1997,ms1998,hs2004,ak2004,bd2012,fmmn2015,ah2015,franco2015,cdk2016,dlp2016,suematsu2016,hty2016,bbgkr2016,cpr2016,ld2017,bck2017,df2017} and the neutrino mass generation have a same scale.

There is another possibility that the neutrino mass generation and the leptogenesis can work at different scales. This scenario has been realized in a tree-level double type-II seesaw model \cite{ghsz2009} and a three-loop neutrino mass model \cite{gu2017}. Similarly we even can construct the models for a high-scale baryogenesis and a low-scale neutron-antineutron oscillation \cite{gs2017}.

In this paper we shall connect a high-scale leptogenesis to a low-scale one-loop neutrino mass generation. In our models, which are based on an $SU(3)_c^{}\times SU(2)_L^{}\times U(1)_Y^{}\times U(1)_{B-L}^{}$ gauge symmetry and a $Z_2^{}$ discrete symmetry, the decays of some $Z_2^{}$-odd real scalar singlets/triplets can produce an asymmetry stored in a $Z_2^{}$-odd scalar doublet. Subsequently, the $Z_2^{}$-odd scalar doublet can decay into the SM lepton doublets and three $Z_2^{}$-odd right-handed neutrinos. Thanks to the sphaleron processes, the lepton asymmetry in the SM leptons can be partially converted to a baryon asymmetry. In this leptogenesis scenario, the $U(1)_{B-L}^{}$ symmetry should be spontaneously broken at a scale below the mass of the $Z_2^{}$-odd scalar doublet. Otherwise, the produced asymmetry in the $Z_2^{}$-odd scalar doublet will be washed out before its conversion to the lepton asymmetry. By integrating out the heavy scalar singlets/triplets, an effective theory can be available for generating the radiative Majorana neutrino masses\cite{zee1980,zee1985,babu1988,bl2001,knt2003,cs2004,ma2006,kms2006,cgn2006,cgnw2006,gs2008,ms2009,lln2009,cg2010,kss2011,chrt2011,cl2012,dabsw2012,kpr2012,ghl2012,lm2013,gnr2013,cght2013,acmn2014,amn2014,gnr2014,ght2014,chao2015,addh2015,cgms2015,wh2015,jtz2015,ckp2015,oo2015,abmy2015,iyz2016,abn2016,dhlx2016,cs2016,amnp2016,noo2016,asw2016,lg2015,no2016,ads2016,gt2016,gms2016,lg2016,lg2016-2,boo2017,koy2017,no2017,blz2017,cos2016,no2017-2,no2017-3,no2017-4,chsvv2017,ccmyz2017} at the TeV scale. Furthermore, the lightest $Z_2^{}$-odd right-handed neutrino can be a dark matter candidate.

\section{The model}

We define the non-SM fermions and scalars as below,
\begin{eqnarray}
\label{nsmf}
&&N_{R}^{}(1,1,0,-1)\,;\nonumber\\
[2mm]
&&\chi(1,1,0,0)=\chi^\ast_{}\,,\nonumber\\
[2mm]
&&\Sigma(1,3,0,0)=\left[\begin{array}{cc} \Sigma^0_{}/\sqrt{2} & \Sigma^+_{} \\
[2mm] \Sigma^-_{} & -\Sigma^0_{}/\sqrt{2} \end{array}\right]=\Sigma^\dagger_{}\,;\nonumber\\
[2mm]&&
\eta(\!\!\begin{array}{c}1,2,-\frac{1}{2},0\end{array}\!\!)=\left[\begin{array}{c} \eta^0_{} \\
[2mm] \eta^-_{} \end{array}\right]=\left[\begin{array}{c} \frac{1}{\sqrt{2}}(\eta^0_{R}+i\eta^0_{I}) \\
[2mm] \eta^-_{} \end{array}\right]\,;\nonumber\\
[2mm]
&&\xi(1,1,0,+2)\,.
\end{eqnarray}
Here and thereafter the brackets following the fields describe the transformations under the $SU(3)_c^{}\times SU(2)_L^{}\times U(1)^{}_{Y}\times U(1)_{B-L}^{}$ gauge groups.

Our models also contain a $Z_2^{}$ discrete symmetry under which only the fields $N^{}_R$, $\chi/\Sigma$ and $\eta$ are odd while the other fields are even, i.e.
\begin{eqnarray}
&&(N_R^{},~\chi/\Sigma,~\eta) \stackrel{Z_2^{}}{\longrightarrow}-(N_R^{},~\chi/\Sigma,~\eta)\,.
\end{eqnarray}
In addition, we introduce a global symmetry $U(1)_G^{}$ under which the fields $N^{}_R$, $\eta$ and $\xi$ respectively carry the numbers $-1$, $+1$, and $+2$, while the other fields are trivial.

We require the $Z_2^{}$ symmetry to be exactly conserved, while the $U(1)_G^{}$ symmetry to be softly broken. The $Z_2^{}$ symmetry will not be broken at any scales and hence the $Z_2^{}$-odd scalars $\chi/\Sigma$ and $\eta$ will not develop any nonzero vacuum expectation values (VEVs). This $Z_2^{}$ symmetry will also forbid the gauge-invariant Yukawa couplings of the $Z_2^{}$-odd right-handed neutrinos $N_R^{}$ to the SM lepton and Higgs doublets. In this sense, we will refer to these $Z_2^{}$-odd fields as the inert fermion singlets, the inert Higgs singlets/triplets and the inert Higgs doublet, respectively.

With all the considerations stated, our models can be classified in the following ways:
\begin{itemize}
\item the model with the inert Higgs singlets other than the inert Higgs triplets,
\begin{eqnarray}
\label{ssd}
\mathcal{L}&\supset&-\rho_\chi^{}\eta^\dagger_{}\chi \phi -y_N^{}\bar{l}_{L}^{}N_{R}^{}\eta-\frac{1}{2}f_N^{}\xi \bar{N}_{R}^{c}N_{R}^{}+\textrm{H.c.}\nonumber\\
&&-\frac{1}{2}M^2_\chi \chi^2_{}-M_{\eta}^2 \eta^\dagger_{}\eta\,,
\end{eqnarray}
\item the model with the inert Higgs triplets other than the inert Higgs singlets,
\begin{eqnarray}
\label{tsd}
\mathcal{L}&\supset&-\sqrt{2}\rho_\Sigma^{} \eta^\dagger_{}\Sigma \phi
-y_N^{}\bar{l}_{L}^{}N_R^{}\eta-\frac{1}{2}f_N^{}\xi\bar{N}_{R}^{c}N_R^{}
+\textrm{H.c.}\nonumber\\
&&-\frac{1}{2}M^2_\Sigma \textrm{Tr}(\Sigma_{}^{2})-M_{\eta}^2 \eta^\dagger_{}\eta\,.
\end{eqnarray}
\item the model with both of the inert Higgs singlets and triplets.
\begin{eqnarray}
\label{sstd}
\mathcal{L}&\supset&-\rho_\chi^{}\eta^\dagger_{}\chi \phi -\sqrt{2}\rho_\Sigma^{} \eta^\dagger_{}\Sigma \phi -y_N^{}\bar{l}_{L}^{}N_{R}^{}\eta\nonumber\\
&&-\frac{1}{2}f_N^{}\xi \bar{N}_{R}^{c}N_{R}^{}+\textrm{H.c.}-\frac{1}{2}M^2_\chi \chi^2_{}-\frac{1}{2}M^2_\Sigma \textrm{Tr}(\Sigma_{}^{2})\nonumber\\
&&-M_{\eta}^2 \eta^\dagger_{}\eta\,.
\end{eqnarray}
\end{itemize}
 Here $\phi$ and $l_{L}^{}$ are the SM Higgs and lepton doublets,
\begin{eqnarray}
\phi(\!\!\begin{array}{c}1,2,-\frac{1}{2},0\end{array}\!\!)=\left[\begin{array}{c} \phi^{0}_{} \\
[2mm] \phi^{-}_{}\end{array}\right]\,,~~l_{L}^{}(\!\!\begin{array}{c}1,2,-\frac{1}{2},-1\end{array}\!\!)=\left[\begin{array}{c} \nu^{}_{L} \\
[2mm] e_{L}^{}\end{array}\right]\,.
\end{eqnarray}

We would like to emphasize that because of the softly broken $U(1)_G^{}$ symmetry, the following quartic term between the SM and inert Higgs doublets is absent from Lagrangians (\ref{ssd}-\ref{sstd}) \cite{lg2016}, i.e.
\begin{eqnarray}
\mathcal{L}\supset\!\!\!\!\!\!/~-\lambda^{}_{}(\eta^\dagger_{}\phi)+\textrm{H.c.}\,.
\end{eqnarray}

\section{Neutrino mass and dark matter}

If the inert Higgs doublet $\eta$ and the inert fermion singlets $N_R^{}$ are much lighter than the inert Higgs singlets/triplets $\chi/\Sigma$, we can integrate out the heavy $\chi/\Sigma$ to induce the following effective quartic coupling between the inert Higgs doublet $\eta$ and the SM Higgs doublet $\phi$, i.e. 
\begin{eqnarray}
\label{eff}
\mathcal{L}\supset-\lambda^{}_{eff}(\eta^\dagger_{}\phi)^2_{}+\textrm{H.c.}~~\textrm{with}~~\lambda_{eff}^{}=-\frac{\rho^2_{\chi/\Sigma}}{M_{\chi/\Sigma}^2}
\end{eqnarray}
In this effective theory, our models can become the Ma model \cite{ma2006} after the Higgs singlet $\xi$ develops its VEV $\langle\xi\rangle \gtrsim 2 \,\textrm{TeV}$ \cite{cddt2004,bbms2009} for spontaneously breaking the $U(1)_{B-L}^{}$ symmetry and then the inert fermion singlets obtain their Majorana masses.

Without loss of generality and for convenience, we can choose the basis where the Majorana mass matrix of the inert fermion singlets $N_R^{}$ is real and diagonal, i.e.
\begin{eqnarray}
M_N^{}=f_N^{}\langle\xi\rangle=\textrm{diag}\{M_{N_1^{}}^{},~M_{N_2^{}}^{},~M_{N_3^{}}^{}\}\,.
\end{eqnarray}
Accordingly we can define the Majorana fermions as below,
\begin{eqnarray}
N_i^{}=N_{Ri}^{}+N_{Ri}^c=N_i^c\,.
\end{eqnarray}

\subsection{Neutrino mass}

\begin{figure*}
\vspace{8cm} \epsfig{file=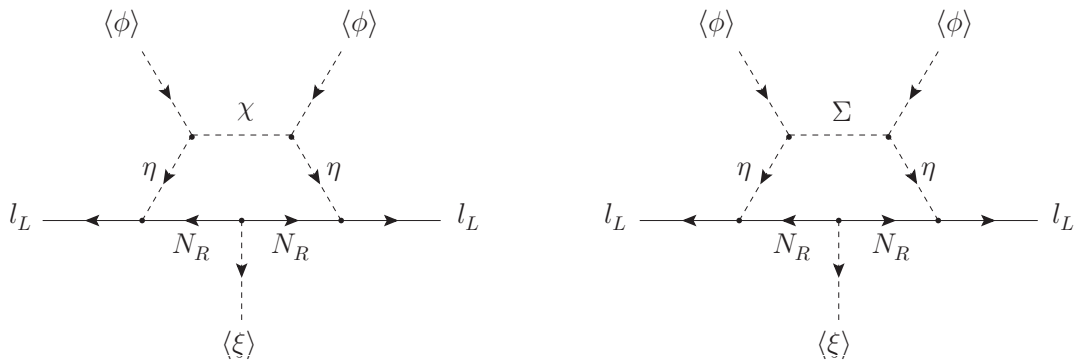, bbllx=6.5cm, bblly=6.0cm,
bburx=16.5cm, bbury=16cm, width=7.5cm, height=7.5cm, angle=0,
clip=0} \vspace{-9cm} \caption{\label{numass} The one-loop neutrino mass generation.}
\end{figure*}

As shown in Fig. \ref{numass}, the neutrinos can obtain a Majorana mass term at one-loop level,
\begin{eqnarray}
\mathcal{L}\supset  -\frac{1}{2}\bar{\nu}_L^{} m_\nu^{} \nu_L^c +\textrm{H.c.}\,.
\end{eqnarray}
The radiative neutrino masses have been given by \cite{ma2006},
\begin{eqnarray}
\label{nmass}
(m_\nu^{})_{\alpha\beta}&=& \frac{1} {16 \pi^2_{}}y_{N\alpha i}^{} M_{N_i}^{} y_{N\beta i}^{} \left[ \frac{M_{\eta^0_R}^2 }{M_{N_i}^2 - M_{\eta^0_R}^2} \ln \left(\frac{M_{N_i}^2}{M_{\eta^0_R}^2}\right)\right.\nonumber\\
&&\left.- \frac{M_{\eta^0_I}^2 }{M_{N_i}^2 - M_{\eta^0_I}^2} \ln \left(\frac{M_{N_i}^2}{M_{\eta^0_I}^2}\right)\right].
\end{eqnarray}
Here $M_{\eta^0_R}^{}$ and $M_{\eta^0_I}^{}$ are the masses of the real and imaginary parts of the neutral component $\eta^0_{}$ of the inert Higgs doublet $\eta$. Clearly, the effective coupling (\ref{eff}) will induce a difference between $M_{\eta^0_R}^{2}$ and $M_{\eta^0_I}^{2}$, i.e.
\begin{eqnarray}
M_{\eta^0_R}^2 - M_{\eta^0_I}^2 =  \lambda_{eff}\langle\phi\rangle^2_{}\,.
\end{eqnarray}
This guarantees the neutrino mass (\ref{nmass}) can arrive at a nonzero value.

The radiative neutrino mass (\ref{nmass}) can be simplified in some limiting cases. For example, we read 
\begin{eqnarray}
\label{nmass2}
\!\!m_\nu^{}\!&=&\! \frac{1} {16 \pi^2_{}} \frac{\lambda_{eff}^{}\langle \phi\rangle^2_{}}{M_{\eta}^2}y^{}_N M_{N}^{} y^T_{N} ~\textrm{for}~M^2_{\eta}\gg \lambda_{eff}^{}\langle \phi\rangle^2_{}\,,\,M_{N}^{2}\,.\nonumber\\
\!\!&&
\end{eqnarray}
It is easy to see the above neutrino masses can be highly suppressed by a small effective coupling $\lambda_{eff}^{}$, even if the inert fermion singlets and the inert Higgs doublet are at the TeV scale while their Yukawa couplings with the SM lepton doublets are sizable. For example, we read 
\begin{eqnarray}
m_\nu^{}&=&0.15\,\textrm{eV}\left(\frac{\lambda_{eff}^{}}{2\times 10^{-6}_{}}\right) \left(\frac{10\,\textrm{TeV}}{M_{\eta}^{}}\right)^2_{}\nonumber\\
&&\times \left(\frac{y_N^{}}{0.2}\right)\left(\frac{M_N^{}}{1\,\textrm{TeV}}\right)\left(\frac{y_N^{T}}{0.2}\right)\,.
\end{eqnarray}
We will show later such small  $\lambda_{eff}^{}$ is inspired by a successful leptogenesis.

\subsection{Dark matter}

As shown in the following section, the inert Higgs doublet $\eta$ should be heavier than the inert fermion singlets $N_R^{}$ for our leptogenesis. It is well known that the lightest one of the inert fermion singlets in the Ma model can keep stable to serve as a dark matter particle. Specifically, in the Ma model, this dark matter particle can annihilate into the SM lepton doublets through the mediation of the inert Higgs doublet. To give a right dark matter relic density, the inert Higgs doublet can not be heavier than $350\,\textrm{GeV}$ or so as the Yukawa couplings $y_N^{}$ are in the perturbative regime, i.e. $y_N^{}\lesssim1$ \cite{kms2006}. In the present models, the dark matter relic density may be dominated by the annihilation of the lightest inert fermion into the SM species through the $s$-channel exchange of the $U(1)_{B-L}^{}$ Higgs boson \cite{os2010}. We hence can relax the constraints on the related masses and couplings.

\section{Leptogenesis}

\begin{figure*}
\vspace{5cm} \epsfig{file=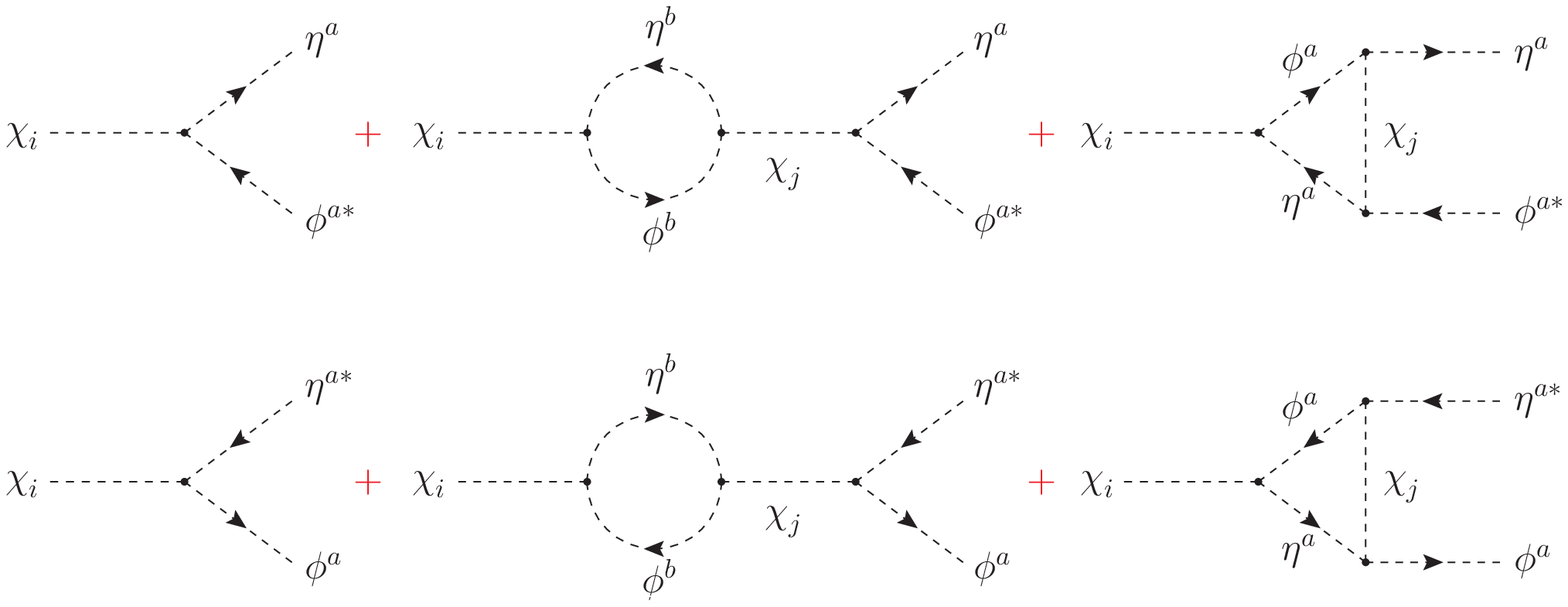, bbllx=6.5cm, bblly=6.0cm,
bburx=16.5cm, bbury=16cm, width=5cm, height=5cm, angle=0,
clip=0} \vspace{-5cm} \caption{\label{hsdecay} The decays of the inert Higgs singlets at one-loop level.}
\end{figure*}

\begin{figure*}
\vspace{6cm} \epsfig{file=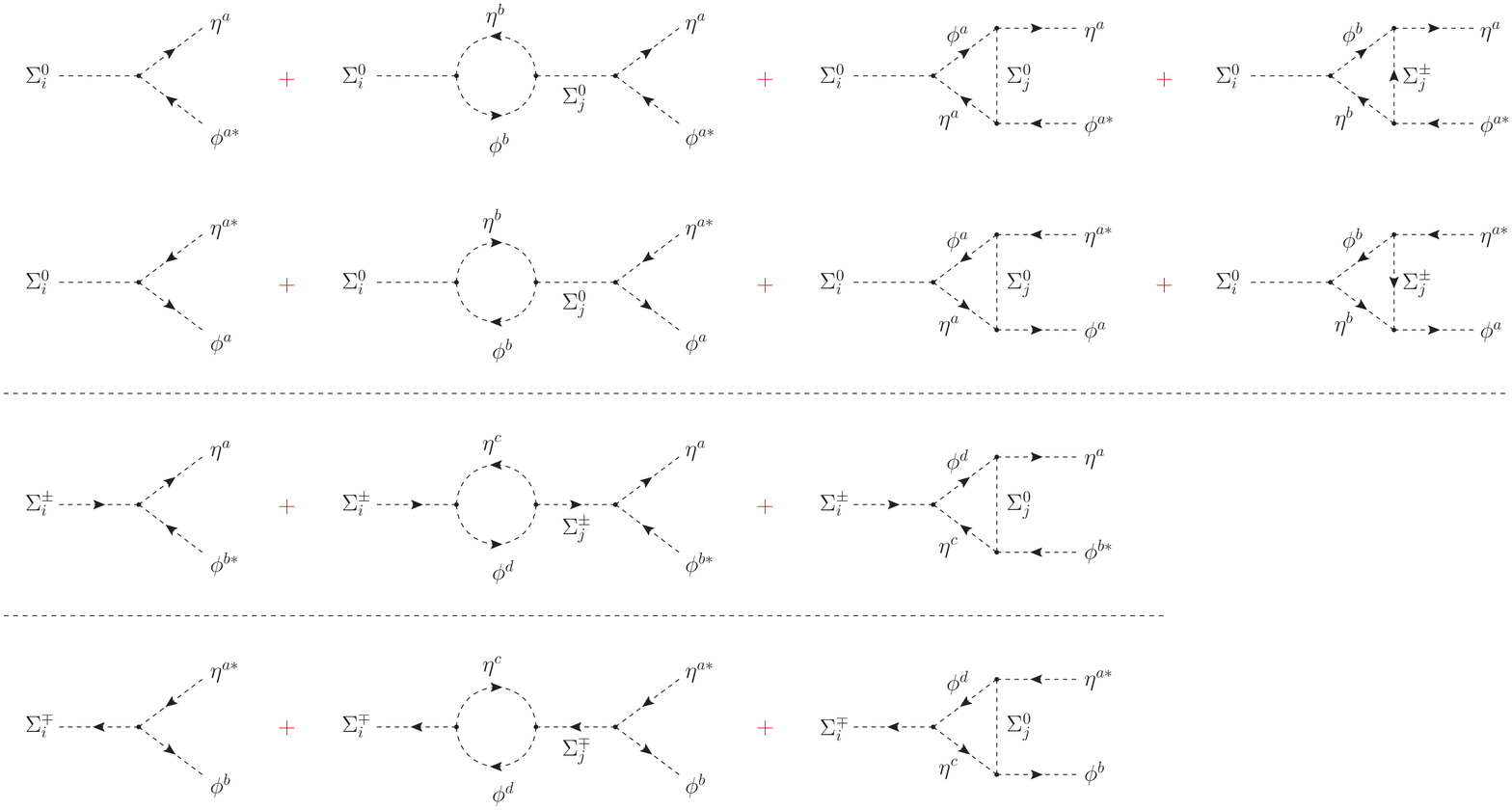, bbllx=13.5cm, bblly=6.0cm,
bburx=23.5cm, bbury=16cm, width=5cm, height=5cm, angle=0,
clip=0} \vspace{0cm} \caption{\label{htdecay} The decays of the inert Higgs triplets at one-loop level.}
\end{figure*}

\begin{figure*}
\vspace{5cm} \epsfig{file=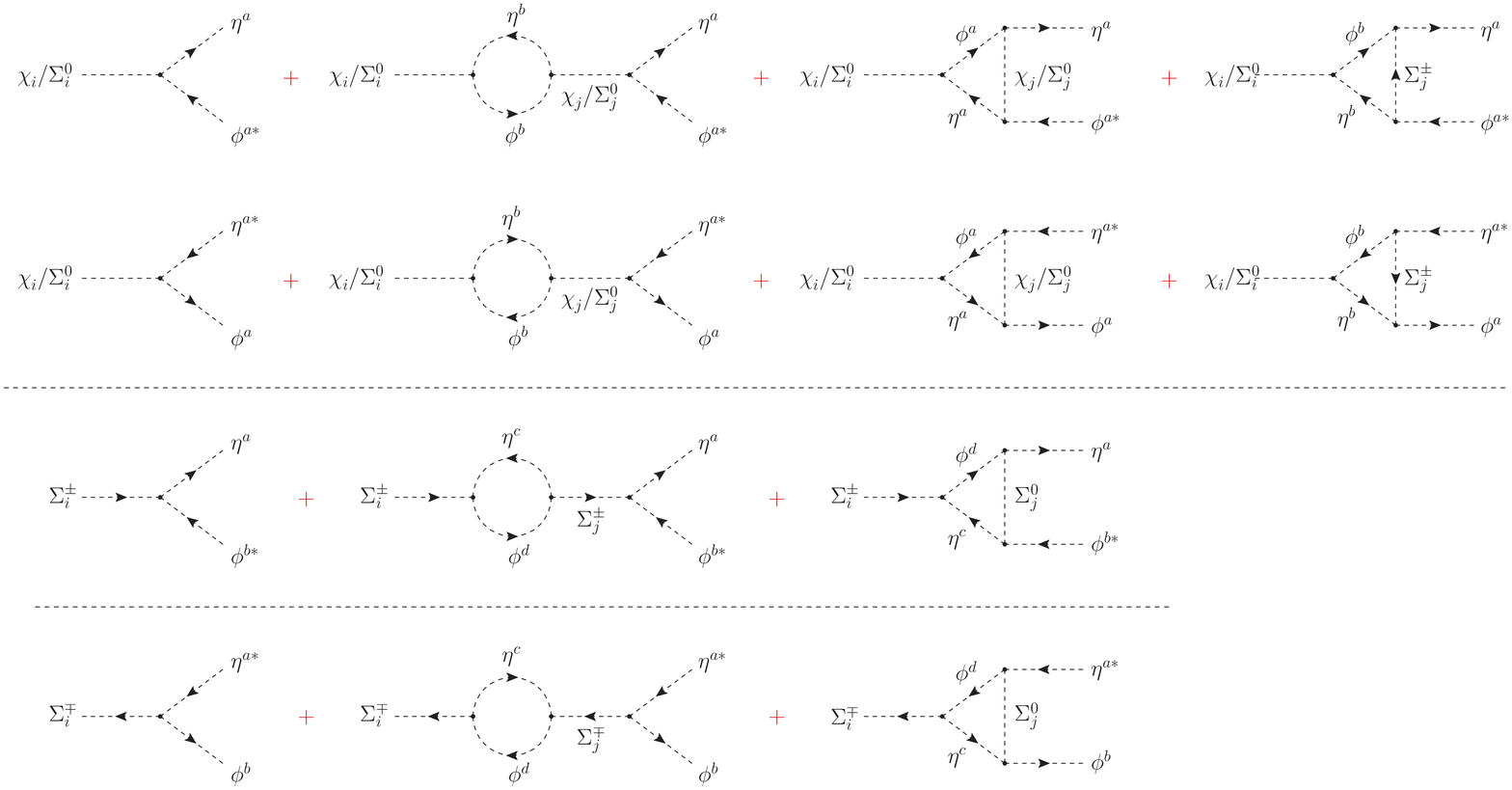, bbllx=13.5cm, bblly=6.0cm,
bburx=23.5cm, bbury=16cm, width=5cm, height=5cm, angle=0,
clip=0} \vspace{0cm} \caption{\label{hstdecay} The decays of the inert Higgs singlets and triplets at one-loop level.}
\end{figure*}

In the Ma model \cite{ma2006}, we can realize a leptogenesis through the CP-violating decays of the inert fermion singlets into the inert Higgs doublets and the SM lepton doublets if these inert fermions are heavy enough. In this leptogenesis scenario, although the inert fermion singlets are allowed at the TeV scale, their Yukawa couplings should be very small for departure from equilibrium. Therefore, it seems difficult for the Ma model to simultaneously offer a successful leptogenesis and a testable neutrino mass generation.

The present models provide another possibility from the decays of the inert Higgs singlets/triplets into the inert Higgs doublets and the SM Higgs doublet. The produced asymmetry stored in the inert Higgs doublet will eventually turn into a lepton asymmetry stored in the SM leptons after the inert Higgs doublet decays into the SM lepton doublets and the inert fermion singlets. The lepton asymmetry in the SM lepton doublets can survive and hence participate in the sphaleron processes if the spontaneous $U(1)_{B-L}^{}$ symmetry breaking scale is not above the inert Higgs doublet mass. We will focus on this leptogenesis scenario and illustrate its main aspects in the following.

\subsection{The inert Higgs singlets/triplets decays}

The decays of the inert Higgs singlets and/or triplets  are shown in Figs. \ref{hsdecay}, \ref{htdecay} and \ref{hstdecay}. We first calculate the decay width at tree level,
\begin{eqnarray}
\label{width}
\Gamma_{S_i^{}}^{}&\equiv &\Gamma(S_i^{}\longrightarrow \eta+\phi^\ast_{})+\Gamma(S_i^{}\longrightarrow \eta^\ast_{}+\phi)\nonumber\\
&=&\frac{1}{4\pi}\frac{|\rho_{S_i^{}}^{}|^2_{}}{M_{S_i^{}}^{}}\,.
\end{eqnarray}
In the above and following formula the field $S_i^{}$ denotes an inert Higgs singlet $\chi_i^{}$ or an inert Higgs triplet $\Sigma_i^{}$.

We then compute the CP asymmetries in the decays of the inert Higgs singlets and/or triplets. At one-loop level, we can obtain these CP asymmetries, 
\begin{eqnarray}
\label{cp1}
\varepsilon_{S_i^{}}^{}&\equiv&\frac{\Gamma(S_i^{}\longrightarrow \eta+\phi^\ast_{})-\Gamma(S_i^{}\longrightarrow \eta^\ast_{}+\phi)}{\Gamma_{S_i^{}}^{}}\nonumber\\
&=&\frac{1}{4\pi}\sum_{j\neq i}^{}\frac{\textrm{Im}\left(\rho^{\ast}_{S_i^{}} \rho_{S_j^{}}^{}\right)^2_{}}{M_{S_i^{}}^2\left|\rho_{S_i^{}}^{}\right|^2_{}}\nonumber\\
&&\times \left[\frac{1}{1-M_{S_j^{}}^2/M_{S_i^{}}^2}+\ln\left(1-\frac{M_{S_i^{}}^2}{M_{S_j^{}}^2}\right)\right]\,.
\end{eqnarray}
If the decaying inert Higgs singlets/triplets $S_i^{}$ are much lighter than the mediating ones $S_j^{}$, the above CP asymmetries can be simplified to be 
\begin{eqnarray}
\label{cp2}
\varepsilon_{S_i^{}}^{}\simeq-\frac{1}{2\pi}\sum_{j\neq i}^{}\frac{\textrm{Im}\left(\rho^{\ast}_{S_i^{}} \rho_{S_j^{}}^{}\right)^2_{}}{M_{S_j^{}}^2\left|\rho_{S_i^{}}^{}\right|^2_{}}~~\textrm{for}~~M_{S_i^{}}^2\ll M_{S_j^{}}^2\,.
\end{eqnarray}

Note the relative CP phases between the parameters $\rho_{S_i^{}}$ and $\rho_{S_j}^{}$ can not be absorbed by any field redefinitions as long as the models contain two or more inert Higgs singlets/triplets. So, the CP asymmetries (\ref{cp1}) and (\ref{cp2}) can always obtain a nonzero value. As an example, we will simply consider two inert Higgs singlets/triplets with hierarchical masses in the following demonstrations. In this case, we can conveniently take 
 \begin{eqnarray}
\rho_{S_1^{}}^{}=\rho_{S_1^{}}^{\ast}\,,~~\rho_{S_2^{}}^{}\longrightarrow \rho_{S_2^{}}^{}e^{i\delta/2}_{}~~\textrm{with}~~\rho_{S_2^{}}^{}=\rho_{S_2^{}}^{\ast}\,,
\end{eqnarray}
and then 
 \begin{eqnarray}
\label{cp3}
\varepsilon_{S_1^{}}^{}\simeq-\frac{1}{2\pi}\frac{\rho_{S_2^{}}^2}{M_{S_2^{}}^2}\sin\delta~~\textrm{for}~~M_{S_1^{}}^2\ll M_{S_2^{}}^2\,.
\end{eqnarray}

\subsection{The final baryon asymmetry}

For the hierarchical spectrum $M_{S_1^{}}^{}\ll M_{S_2^{}}^{}$, the final baryon asymmetry should be mostly produced by the decays of the lightest inert Higgs singlet/triplet $S_1^{}$. 
Instead of fully integrating the Boltzmann equations to determine the final baryon asymmetry, we shall adopt an instructive and reliable estimation. For this purpose, we define a useful parameter,
\begin{eqnarray}
\label{rwidth}
K&=&\frac{\Gamma_{S_1^{}}^{}}{2H(T)}\left|_{T=M_{S_1^{}}^{}}^{}\right.\,,
\end{eqnarray}
with $H(T)$ being the Hubble constant,
\begin{eqnarray}
H=\left(\frac{8\pi^{3}_{}g_{\ast}^{}}{90}\right)^{\frac{1}{2}}_{}
\frac{T^{2}_{}}{M_{\textrm{Pl}}^{}}\,.
\end{eqnarray}
Here $g_\ast^{}=116$ is the relativistic degrees of freedom during the leptogenesis epoch and $M_{\textrm{Pl}}^{}=1.22\times 10^{19}_{}\,\textrm{GeV}$ is the Planck mass.

In the weak washout region with $K\ll 1$\footnote{Here the Higgs portal interactions are assumed weak enough so that they can decouple before the leptogenesis epoch. For the inert Higgs triplets, their gauge interactions will not significantly affect the leptogenesis if they are heavy enough.}, the final baryon asymmetry can well approximate to \cite{kt1990}
\begin{eqnarray}
\label{basy}
\eta_B^{}&=&\frac{n_B^{}}{s}\simeq -\frac{28}{79}\times \frac{\varepsilon_{S_1^{}}^{}}{g_\ast^{}} \times C~~\textrm{for}~~K\ll 1\,.
\end{eqnarray}
Here $n_B^{}$ and $s$, respectively, are the baryon number density and the entropy density, the number $-\frac{28}{79}$ is the sphaleron lepton-to-baryon coefficient, while the factor $C$ depends on the decaying inert Higgs singlet/triplet, i.e. $C=1$ for $S_{1}^{}=\chi_1^{}$ or $C=3$ for $S_1^{}=\Sigma_1^{}$.

\subsection{The lepton-to-baryon conversion}

We now analyze the chemical potentials \cite{ht1990} to discuss the details of the conversion between the lepton and baryon asymmetries. For this purpose, we denote $\mu_{q,d,u}^{}$ for the chemical potentials of the SM quarks $q_{L}^{}(3,2,+\frac{1}{6},+\frac{1}{3})$, $d_{R}^{}(3,1,-\frac{1}{3},+\frac{1}{3})$ and $u_{R}^{}(3,1,+\frac{2}{3},+\frac{1}{3})$, while $\mu_{l,e,\phi,N, \eta,\xi}^{}$ for the chemical potentials of the SM leptons and Higgs scalar $l_L^{}$, $e_R^{}$, $\phi$ as well as the non-SM fields $N_R^{}$, $\eta$, $\xi$. We further assume the $U(1)_{B-L}^{}$ symmetry will be spontaneously broken at a scale below the inert Higgs doublet mass but before the electroweak symmetry breaking. This choice can escape from the potentially experimental constraints if the $U(1)_{B-L}^{}$ gauge coupling $g_{B-L}^{}$ is small enough. We then can consider the chemical potentials in three phases,
\begin{itemize}
\item phase-1 during the inert Higgs singlets/triplets decays and the inert Higgs doublet decays, 
\item phase-2 during the inert Higgs doublet decays and the $U(1)_{B-L}^{}$ symmetry breaking,
\item phase-3 during the $U(1)_{B-L}^{}$ symmetry breaking and the electroweak symmetry breaking.
\end{itemize}

In phase-1, the SM Yukawa interactions are in equilibrium and hence yield \cite{ht1990},
\begin{eqnarray}
\label{chemical1}
-\mu_{q}^{}-\mu_{\phi}^{}+\mu_{d}^{}&=&0\,,\nonumber\\
 -\mu_{q}^{}+\mu_{\phi}^{}+\mu_{u}^{}&=&0\,,\nonumber\\
 -\mu_{l}^{}-\mu_{\phi}^{}+\mu_{e}^{}&=&0\,,
\end{eqnarray}
the fast sphalerons constrain \cite{ht1990},
\begin{eqnarray}
\label{chemical2}
3\mu_{q}^{}+\mu_{l}^{}&=&0\,,
\end{eqnarray}
while the neutral hypercharge in the universe requires,
\begin{eqnarray}
\label{chemical3}
3\left( \mu_{q}^{} -\mu_{d}^{}+2\mu_{u}^{}-\mu_{l}^{} -\mu_{e}^{}\right)-2\mu_{\phi}^{}- 2\mu_{\eta}^{} =0\,.
\end{eqnarray}
In addition, the non-SM Yukawa interactions in Eqs. (\ref{ssd}-\ref{sstd}) are also in equilibrium. This means
\begin{eqnarray}
\label{chemical4}
-\mu_{l}^{}+\mu_{N}^{}+\mu_{\eta}^{}=0\,,~~\mu_{\xi}^{}+2\mu_{N}^{}=0\,.
\end{eqnarray}
Furthermore, we should consider the effective coupling (\ref{eff}). We can roughly estimate the interaction rate $\Gamma \sim \lambda_{eff}^2 T /(4\pi) < H(T)$ and hence this effective coupling could keep in equilibrium during the temperatures $ M_{\eta}^{}\lesssim T \lesssim 300\, \lambda_{eff}^2 M_{\textrm{Pl}}^{}$. So, we have 
\begin{eqnarray}
\label{chemical5}
2\mu_{\phi}^{}-2\mu_{\eta}^{}=0\,.
\end{eqnarray}
Note in Eqs. (\ref{chemical1}-\ref{chemical4}), we have identified the chemical potentials of the different-generation fermions because the Yukawa interactions establish an equilibrium between the different generations. By solving Eqs. (\ref{chemical1}-\ref{chemical5}), we find
\begin{eqnarray}
\label{chemical6}
&&\mu_q^{}=-\frac{1}{3}\mu_l^{}\,,~~\mu_d^{}=-\frac{5}{6}\mu_l^{}\,,~~\mu_u^{}=\frac{1}{6}\mu_l^{}\,,~~\mu_{e}^{}=\frac{1}{2}\mu_l^{}\,,\nonumber\\
&&\mu_\phi^{}=\mu_\eta^{}=-\frac{1}{2}\mu_l^{}\,,~~\mu_{N}^{}=\frac{3}{2}\mu_l^{}\,,~~\mu_\xi^{}=-3\mu_l^{}\,.
\end{eqnarray}
Clearly, if the $U(1)_{B-L}^{}$ symmetry was broken in this stage, the inert fermion singlets $N_R^{}$ would have a Majorana mass term and hence they can not have any nonzero chemical potentials. Accordingly, the chemical potentials of other species would also arrive at a zero value. This is because the effective coupling (\ref{eff}) will go into equilibrium after it keeps departure from equilibrium for some time\footnote{Similar to the effective coupling (\ref{eff}),  an effective coupling between two charged scalars exists in the models of Refs. \cite{gu2017} and \cite{gs2017}. To revive the high-scale leptogenesis/baryogenesis in those models, we can introduce a spontaneous symmetry breaking to generate the related Majorana masses before the charged scalar decays.}.

In phase-2, the inert Higgs doublet $\eta$ has already decayed so that the condition (\ref{chemical3}) for the zero hypercharge should be modified by \cite{ht1990}, 
\begin{eqnarray}
\label{chemical7}
3\left(\mu_{q}^{}-\mu_{d}^{}+2\mu_{u}^{}- \mu_{l}^{}-\mu_{e}^{}\right)-2\mu_{\phi}^{}=0\,.
\end{eqnarray}
Ones then can solve Eqs. (\ref{chemical1}), (\ref{chemical2}), (\ref{chemical4}) and (\ref{chemical7}) to determine the chemical potentials,
\begin{eqnarray}
\label{chemical8}
&&\mu_q^{}=-\frac{1}{3}\mu_l^{}\,,~~\mu_d^{}=-\frac{19}{21}\mu_l^{}\,,~~\mu_u^{}=\frac{5}{21}\mu_l^{}\,,~~\mu_{e}^{}=\frac{3}{7}\mu_l^{}\,,\nonumber\\
&&\mu_\phi^{}=-\frac{4}{7}\mu_l^{}\,,~~~~\mu_{N}^{}=-\frac{1}{2}\mu_\xi^{}\,.
\end{eqnarray}

In phase-3, the Higgs singlet $\xi$ develops its VEV while the inert fermion singlets $N_R^{}$ acquire their Majorana masses. Accordingly, Eq. (\ref{chemical4}) now should be
\begin{eqnarray}
\label{chemical9}
2\mu_{N}^{}=0\,.
\end{eqnarray}
The nonzero chemical potentials can be given by Eqs. (\ref{chemical1}), (\ref{chemical2}) and (\ref{chemical7}), i.e. \cite{ht1990}
\begin{eqnarray}
\label{chemical10}
&&\mu_q^{}=-\frac{1}{3}\mu_l^{}\,,~~\mu_d^{}=-\frac{19}{21}\mu_l^{}\,,~~\mu_u^{}=\frac{5}{21}\mu_l^{}\,,~~\mu_{e}^{}=\frac{3}{7}\mu_l^{}\,,\nonumber\\
&&\mu_\phi^{}=-\frac{4}{7}\mu_l^{}\,.
\end{eqnarray}
Therefore, the relation between the final baryon and lepton numbers should be \cite{ht1990} 
\begin{eqnarray}
B&=&\frac{28}{79}\left(B-L\right)~~\textrm{with}\nonumber\\
[2mm]
&&B\propto \frac{1}{3}(\mu_q^{} \times 2+ \mu_d^{} + \mu_u^{})\times 3 \times 3 =-4\mu_l^{}\,,\nonumber\\
[2mm]
&&L\propto (\mu_l^{} \times 2+ \mu_e^{} )\times 3  =\frac{51}{7}\mu_l^{}\,.
\end{eqnarray}

\subsection{Numerical estimation}

To provide a numerical illustration, we consider the model with two inert Higgs singlets. After setting the inputs,
\begin{eqnarray}
\label{par1}
&&M_{\chi_1^{}}^{}=10^{3}_{}\rho_{\chi_1^{}}^{}=10^{13}_{}\,\textrm{GeV}\,,\nonumber\\
&&M_{\chi_2^{}}^{}=10^{3}_{}\rho_{\chi_2^{}}^{}=10^{14}_{}\,\textrm{GeV}\,,~~\sin\delta =0.2\,,
\end{eqnarray}
in Eqs. (\ref{width}), (\ref{cp2}), (\ref{rwidth}) and (\ref{basy}), we read
\begin{eqnarray}
&&\varepsilon_{\chi_1^{}}^{}=0.032\,,~~K=0.0054\,,
\end{eqnarray}
and then obtain the baryon asymmetry,
\begin{eqnarray}
\eta_B^{}=10^{-10}_{}\,.
\end{eqnarray}

\section{Summary}

In this paper we have demonstrated that a high-scale leptogenesis can be consistent with a low-scale one-loop neutrino mass generation. In our models, the decays of some heavy inert Higgs singlets/triplets can produce an asymmetry stored in an inert Higgs doublet. Subsequently, the inert Higgs doublet can decay into the SM lepton doublets and the inert fermion singlets. Thanks to the sphaleron processes, the lepton asymmetry in the SM leptons can be partially converted to a baryon asymmetry. In this leptogenesis scenario, a $U(1)_{B-L}^{}$ gauge symmetry should be spontaneously broken at a scale below the mass of inert Higgs doublet. Otherwise, the produced inert Higgs doublet asymmetry will be washed out before its conversion to the lepton asymmetry. By integrating out the Higgs singlets/triplets, we can obtain an effective theory to radiatively generate the neutrino masses at the TeV scale. Furthermore, the lightest inert fermion singlet can keep stable to act as a dark matter particle.

\textbf{Acknowledgement}: This work was supported by the National Natural Science Foundation of China under Grant No. 11675100, the Recruitment Program for Young Professionals under Grant No. 15Z127060004, the Shanghai Jiao Tong University under Grant No. WF220407201, the Shanghai Laboratory for Particle Physics and Cosmology, and the Key Laboratory for Particle Physics, Astrophysics and Cosmology, Ministry of Education.


\begin{thebibliography}{16}




\bibitem{patrignani2016}
C. Patrignani {\it et al.}, (Particle Data Group Collaboration), Chin. Phys. C \textbf{40}, 1000001 (2016).


\bibitem{minkowski1977}
P. Minkowski, Phys. Lett. B \textbf{67}, 421 (1977).

\bibitem{yanagida1979}
T. Yanagida, {\it Proceedings of the Workshop on Unified Theory and the Baryon
Number of the Universe}, ed. O. Sawada and A. Sugamoto (Tsukuba 1979).

\bibitem{grs1979}
M. Gell-Mann, P. Ramond, and R. Slansky, {\it Supergravity}, ed. F. van Nieuwenhuizen and D. Freedman
(North Holland 1979).

\bibitem{ms1980}
R.N. Mohapatra and G. Senjanovi\'{c}, Phys.
Rev. Lett. \textbf{44}, 912 (1980).

\bibitem{mw1980}
M. Magg and C. Wetterich, Phys. Lett. B \textbf{94}, 61 (1980).


\bibitem{sv1980}
J. Schechter and J.W.F. Valle, Phys. Rev. D \textbf{22}, 2227 (1980).

\bibitem{cl1980}
T.P. Cheng and L.F. Li, Phys. Rev. D \textbf{22}, 2860 (1980).


\bibitem{lsw1981}
G. Lazarides, Q. Shafi, and C. Wetterich, Nucl. Phys. B \textbf{181},
287 (1981).



\bibitem{ms1981}
R.N. Mohapatra and G. Senjanovi\'{c}, Phys. Rev. D
\textbf{23}, 165 (1981).



\bibitem{flhj1989}
R. Foot, H. Lew, X.G. He, and G.C. Joshi, Z. Phys. C \textbf{44},
441 (1989).


\bibitem{ma1998}
E. Ma, Phys. Rev. Lett. \textbf{81}, 1171 (1998).





\bibitem{fy1986}
M. Fukugita and T. Yanagida, Phys. Lett. B \textbf{174}, 45 (1986).





\bibitem{krs1985}
V.A. Kuzmin, V.A. Rubakov, and M.E. Shaposhnikov, Phys. Lett. B \textbf{155}, 36 (1985).



\bibitem{lpy1986}
P. Langacker, R.D. Peccei, and T. Yanagida, Mod. Phys. Lett. A
\textbf{1}, 541 (1986). 


\bibitem{luty1992}
M.A. Luty, Phys. Rev. D \textbf{45}, 455 (1992).


\bibitem{mz1992}
R.N. Mohapatra and X. Zhang, Phys. Rev. D \textbf{46}, 5331 (1992).








\bibitem{fps1995}
M. Flanz, E.A. Paschos, and U. Sarkar, Phys. Lett. B \textbf{345},
248 (1995). 


\bibitem{crv1996}
L. Covi, E. Roulet, and F. Vissani, Phys. Lett. B \textbf{384}, 169 (1996).



\bibitem{pilaftsis1997}
A. Pilaftsis, Phys. Rev. D \textbf{56}, 5431 (1997).





\bibitem{ms1998}
E. Ma and U. Sarkar, Phys. Rev. Lett. \textbf{80}, 5716 (1998).




\bibitem{hs2004}
T. Hambye and G. Senjanovi\'{c}, Phys. Lett. B \textbf{582}, 73
(2004). 



\bibitem{ak2004}
S. Antusch and S.F. King, Phys. Lett. B \textbf{597}, 199
(2004).


\bibitem{bd2012}
S. Blanchet and P. Di Bari, New J.Phys. \textbf{14}, 125012  (2012).


\bibitem{fmmn2015}
C.S. Fong, D. Meloni, A. Meroni, and E. Nardi, JHEP \textbf{1501}, 11 (2015).

\bibitem{ah2015}
M. Aoki  and N. Haba, PTEP \textbf{2015}, 113B03 (2015).




\bibitem{franco2015}
E.T. Franco, Phys. Rev. D \textbf{92}, 113010 (2015).




\bibitem{cdk2016}
P. Chen, G.J. Ding, and S.F. King, JHEP \textbf{1603}, 206 (2016).

\bibitem{dlp2016}
P. Di Bari, P. O. Ludl, and S. Palomares-Ruiz, JCAP \textbf{1611}, 044 (2016).


\bibitem{suematsu2016}
D. Suematsu, Phys. Lett. B \textbf{760}, 538 (2016).


\bibitem{hty2016}
Y. Hamada, K. Tsumura, and D. Yasuhara, Phys. Rev. D \textbf{95},103505 (2017). 





\bibitem{bbgkr2016} 
G. Bambhaniya, P.?S. Bhupal Dev, S. Goswami, S. Khan, and W. Rodejohann, Phys. Rev. D \textbf{95}, 095016 (2017).




\bibitem{cpr2016}
B. Charles Bryant, Z. Poh, and S. Raby, arXiv:1612.04382 [hep-ph].

\bibitem{ld2017}	
 C.C. Li, G.J. Ding, arXiv:1701.08508 [hep-ph].

\bibitem{bck2017}
A. Biswas, S. Choubey, and S. Khan, arXiv:1704.00819 [hep-ph].


\bibitem{df2017} 
P. Di Bari and M. Re Fiorentin, arXiv:1705.01935 [hep-ph].








  
  
  

  
  
  
\bibitem{ghsz2009} 
  P.H. Gu, H.J. He, U. Sarkar, and X. Zhang, Phys. Rev. D \textbf{80}, 053004 (2009).  
  
  
\bibitem{gu2017}  
P.H. Gu, JHEP \textbf{1704}, 159 (2017).



  
\bibitem{gs2017}
P.H. Gu and U. Sarkar, arXiv:1705.02858 [hep-ph].









\bibitem{zee1980}
A. Zee, Phys. Lett. B \textbf{93}, 389 (1980).




\bibitem{zee1985}
A. Zee, Phys. Lett. B \textbf{161}, 141 (1985).

\bibitem{babu1988}
K. S. Babu, Phys. Lett. B \textbf{203}, 132 (1988).

\bibitem{bl2001}
K.S. Babu and C.N. Leung, Nucl. Phys. B 619, 667
(2001).


\bibitem{knt2003}
L.M. Krauss, S. Nasri, and M. Trodden, Phys. Rev. D \textbf{67}, 085002 (2003).

\bibitem{cs2004}
K. Cheung and O. Seto, Phys. Rev. D \textbf{69},
113009 (2004).




\bibitem{ma2006}
E. Ma, Phys. Rev. D \textbf{73}, 077301 (2006).



\bibitem{kms2006}
J. Kubo, E. Ma, and D. Suematsu, Phys. Lett. B \textbf{642}, 18 (2006).


\bibitem{cgn2006}
C.S. Chen, C.Q. Geng, and J.N. Ng, Phys. Rev. D \textbf{75}, 053004 (2007).


\bibitem{cgnw2006}
C.S. Chen, C.Q. Geng, J.N. Ng, and J.M.S Wu, JHEP \textbf{0708}, 022 (2007).



\bibitem{gs2008}
P.H. Gu and U. Sarkar, Phys. Rev. D \textbf{78}, 073012 (2008).




\bibitem{ms2009}
E. Ma and D. Suematsu, Mod. Phys. Lett. A \textbf{24}, 583 (2009).



\bibitem{lln2009}
Y. Liao and J.Y. Liu, and G.Z. Ning, Phys. Rev. D \textbf{79}, 073003 (2009).




\bibitem{cg2010}	
C.S. Chen and C.Q. Geng, Phys. Rev. D \textbf{82}, 105004 (2010).




\bibitem{kss2011}
S. Kanemura, O. Seto, and T. Shimomura, Phys. Rev. D \textbf{84}, 016004 (2011).



\bibitem{chrt2011}
Y. Cai, X.G. He, M. Ramsey-Musolf, L.H. Tsai, JHEP \textrm{1112}, 054 (2011).

\bibitem{cl2012}
C.H. Chen and S.S.C. Law, Phys. Rev. D \textbf{85}, 055012 (2012).






\bibitem{dabsw2012}
F. del Aguila, A. Aparici, S. Bhattacharya, A. Santamaria, and J. Wudka, JHEP \textbf{1205}, 133 (2012).

\bibitem{kpr2012}
K. Kumericki, I. Picek, and B. Radovcic, JHEP \textbf{1207}, 039 (2012).



\bibitem{ghl2012}
G. Guo, X.G. He, and G.N. Li, JHEP \textbf{1210}, 044 (2012).




\bibitem{lm2013}
S.S.C. Law and K.L. McDonald, JHEP \textbf{1309}, 092 (2013).



\bibitem{gnr2013}
M. Gustafsson, J.M. No, and M.A. Rivera, Phys. Rev. Lett. \textbf{110}, 211802 (2013).



\bibitem{cght2013}
C.S. Chen, C.Q. Geng, D. Huang, and L.H. Tsai, Phys. Rev. D \textbf{87}, 077702 (2013).




\bibitem{acmn2014}
A. Ahriche, C.S. Chen, K.L. McDonald, and S. Nasri, Phys. Rev. D \textbf{90}, 015024
(2014).



\bibitem{amn2014}
A. Ahriche, K. L. McDonald, and S. Nasri, JHEP \textbf{1410}, 167 (2014).



\bibitem{gnr2014}
M. Gustafsson, J.M. No, and M.A. Rivera,  Phys.Rev. D \textbf{90}, 013012 (2014).

\bibitem{ght2014} 
C.Q. Geng, D. Huang, L.H. Tsai, Phys. Rev. D \textbf{90}, 113005 (2014).

	
	
	
	

\bibitem{chao2015}
W. Chao, Int. J. Mod. Phys. A \textbf{30}, 1550007 (2015).

\bibitem{addh2015}
D. Aristizabal Sierra, A. Degee, L. Dorame, and M. Hirsch, JHEP \textbf{1503}, 040 (2015).

\bibitem{cgms2015}
N. Chakrabarty, D.K. Ghosh, B. Mukhopadhyaya, and I. Saha, Phys. Rev. D \textbf{92}, 015002 (2015).



\bibitem{wh2015}
 W. Wang and Z.L. Han, Phys.Rev. D \textbf{92}, 095001 (2015).


\bibitem{jtz2015}
L.G. Jin, R. Tang, and F. Zhang, Phys. Lett. B \textbf{741}, 163 (2015).

\bibitem{ckp2015}
P. Culjak, K. Kumericki, and I. Picek, Phys. Lett. B \textbf{744}, 237 (2015).

\bibitem{oo2015}
H. Okada and Y. Orikasa, Phys.Rev. D \textbf{94}, 055002 (2016).

\bibitem{abmy2015}
A. Arhrib, C. B{\oe}hm, E. Ma, and T.C. Yuan, JCAP \textbf{1604}, 049 (2016).

\bibitem{iyz2016}
A. Ibarra, C.E. Yaguna, and O. Zapata, Phys. Rev. D \textbf{93}, 035012 (2016).

\bibitem{abn2016}
A. Ahriche, S.M. Boucenna, and S. Nasri, Phys. Rev. D \textbf{93}, 075036 (2016). 



\bibitem{dhlx2016}
R. Ding, Z.L. Han, Y. Liao, and W.P. Xie, JHEP \textbf{1605}, 030 (2016).

\bibitem{cs2016}
Y. Cai and M.A. Schmidt, JHEP \textbf{1605}, 028 (2016).

\bibitem{amnp2016}
A. Ahriche, K.L. McDonald, S. Nasri, and I. Picek, Phys. Lett. B \textbf{757}, 399 (2016).


\bibitem{noo2016}
T. Nomura, H. Okada, and Y. Orikasa, Phys. Rev. D \textbf{93}, 113008 (2016).


\bibitem{asw2016}
D. Aristizabal Sierra, C. Simoes, and D. Wegman, JHEP \textbf{1606}, 108 (2016).



\bibitem{lg2015}
W.B. Lu and P.H. Gu, JCAP \textbf{1605}, 040 (2016).



\bibitem{no2016}
T. Nomura and H. Okada, Phys. Rev. D \textbf{94}, 093006 (2016).



\bibitem{ads2016}
J. Alcaide, D. Das, A. Santamaria, JHEP \textbf{1704}, 049 (2017).



\bibitem{gt2016}	
C.Q. Geng and L.H. Tsai, Annals Phys. \textbf{365}, 210 (2016).


\bibitem{gms2016} 
P.H. Gu, E. Ma, and U. Sarkar, Phys. Rev. D \textbf{94}, 111701 (2016).



\bibitem{lg2016}
Z. Liu and P.H. Gu, Nucl. Phys. B \textbf{915}, 206 (2017).

\bibitem{lg2016-2}
W.B. Lu and P.H. Gu, arXiv:1611.02106 [hep-ph].



\bibitem{boo2017}
S. Baek, H. Okada, and Y. Orikasa, arXiv:1703.00685 [hep-ph].




\bibitem{koy2017}
T. Kitabayashi, S. Ohkawa, M. Yasu\'{e}, arXiv:1703.09417 [hep-ph].



\bibitem{no2017}
T. Nomura and H. Okada, arXiv:1704.08581 [hep-ph].



\bibitem{blz2017}
A. Betancur, R. Longas, and O. Zapata, arXiv:1704.01162 [hep-ph].



 \bibitem{cos2016} 
C.W. Chiang, H. Okada, and E. Senaha, Phys. Rev. D \textbf{96}, 015002 (2017).




\bibitem{no2017-2}
T. Nomura and H. Okada, Phys. Rev. D \textbf{96}, 015016 (2017).


 
 
 
 
 
 
 
 \bibitem{no2017-3} 
 T. Nomura and H. Okada, arXiv:1705.08309 [hep-ph].



 \bibitem{no2017-4} 
 T. Nomura and H. Okada, arXiv:1706.05268 [hep-ph].
 
 
 


\bibitem{chsvv2017}
Y. Cai, J. Herrero-Garc\'{i}a, M.A. Schmidt, A. Vicente, and R.R. Volkas, arXiv:1706.08524 [hep-ph].


\bibitem{ccmyz2017}
Q.H. Cao, S.L. Chen, E. Ma, B. Yan, D.M. Zhang, arXiv:1707.05896 [hep-ph].



\bibitem{cddt2004}
M.S. Carena, A. Daleo, B.A. Dobrescu, and T.M.P. Tait, Phys. Rev. D \textbf{70}, 093009 (2004).

\bibitem{bbms2009}
L. Basso, A. Belyaev, S. Moretti, and C.H. Shepherd-Themistocleous, Phys. Rev. D \textbf{80}, 055030 (2009).








\bibitem{os2010}
N. Okada and O. Seto, Phys. Rev. D \textbf{82}, 023507 (2010).






\bibitem{kt1990}
E.W. Kolb and M.S. Turner, \textit{The Early Universe},
Addison-Wesley, 1990.

\bibitem{ht1990}
J.A. Harvey and M.S. Turner, Phys. Rev. D \textbf{42}, 3344 (1990).




\end{thebibliography}
\end{document}